\documentclass[aps,pra,onecolumn,groupedaddress,showpacs]{revtex4}

\usepackage{amssymb,amsmath}
\usepackage{graphicx}

\newcommand{\om}{\omega}

\newcommand{\pa}{\partial}

\begin{document}

\title{Two-component nonlinear wave of the Hirota equation}

\author{G. T. Adamashvili}
\affiliation{Technical University of Georgia, Kostava str.77, Tbilisi, 0179, Georgia.\\ email: $guram_{-}adamashvili@ymail.com.$ }

\begin{abstract}
Using the generalized perturbation reduction method the Hirota equation is transformed to the coupled nonlinear Schr\"odinger equations for auxiliary functions. A solution in the form of a two-component vector nonlinear pulse is obtained. The components of the pulse oscillate with the sum and difference of the frequencies and the wave numbers.
Explicit analytical expressions for the shape and parameters of the two-component nonlinear pulse are presented.
$$$$
\end{abstract}


\maketitle

\section{Introduction}

Nonlinear solitary waves are one of the most extensive subjects to research in various nonlinear physical systems. The single-component and two-component nonlinear solitary  waves such as soliton, breather, vector breather are considered. The nonlinear solitary wave behavior can analytically be described in nonlinear partial differential equations. Some of the equations are the Sine-Gordon equation, the nonlinear Schr\"odinger equation, the Maxwell-Bloch equation, the Korteweg de Vries equation, and many others. Various methods in solving the nonlinear wave equations such as the inverse scattering transform, the perturbative reduction method, and many others, have been successfully applied [1-8]. It is necessary to separately consider the nonlinear partial differential equations for real and complex functions. Some of equations for complex function $u$  are the nonlinear Schr\"odinger equation, the complex modified Korteweg-de Vries equation and the Hirota equation among them [9,10]
\begin{equation}\label{hir}
 i u_{ t}+ \rho u_{zz} +\delta |u|^{2} u  +i\sigma u_{zzz}    +i 3 \alpha |u|^{2} u_{z} =0,
\end{equation}
where $u(z,t)$ is a complex valued function of the spatial coordinate $z$ and time $t$. $\alpha,\;\rho,\;\sigma,\; \delta$ are the real constants and for which the condition
\begin{equation}\label{coef}
\alpha \rho=\sigma \delta
\end{equation}
is valid.

When  $\alpha=\sigma=0$,  the Hirota equation (1) can be reduced to the nonlinear Schr\"odinger equation
\begin{equation}\label{nse}\nonumber
 i u_{ t}+ \rho u_{zz} +\delta |u|^{2} u  =0.
\end{equation}
But,  when $\rho=\delta=0$,  Eq.(1) is transformed to the complex modified Korteweg-de Vries equation
\begin{equation}\label{kdv}\nonumber
 u_{ t} +\sigma u_{zzz} + 3 \alpha |u|^{2} u_{z}  =0.
\end{equation}

Sometimes, it is convenient  rewritten the Hirota equation (1) to the following form [2]
\begin{equation}\label{Hir3}\nonumber
 i \frac{\pa u}{\pa t} -\alpha_{2}( \frac{\pa^{2} u}{\pa z^{2}} +2 |u|^{2} u)  +i\alpha_{3}( \frac{\pa^{3} u}{\pa z^{3}}    +6|u|^{2} \frac{\pa u}{\pa z}) =0,
\end{equation}
where
$$
\alpha=2\alpha_{3},\;\;\;\;\;\;\;\;\;\;\;\;\;\delta=-2\alpha_{2},\;\;\;\;\;\;\;\;\;\;\;\;\;\rho=-\alpha_{2},\;\;\;\;\;\;\;\;\;\;\;\;\;\sigma=\alpha_{3},
$$
in such a way that the constraint Eq.(2) is satisfied.

The complete solution of the Hirota equation \eqref{hir} by means of the inverse scattering transform is obtained [2].

We consider the real function $U(z,t)$ of the space coordinate $z$ and time $t$  which can be presented to the form
\begin{equation}\label{U1}
 U=u+u^{*},
\end{equation}
where the complex function $u(z,t)$ is the solution of the Hirota equation (1), $*$ stands for the complex conjugate.

When the duration of the pulse $T >>\omega^{-1}$ we can use the slowly varying envelope approximation [11-14]. In this case we represent the complex conjugation functions $u$ and $u^{*}$ in the following form
\begin{equation}\label{2}
 u=\hat{u}_{+1}Z_{+1},\;\;\;\;\;\;\;\;\;\;\;\; u^{*}=\hat{u}_{-1}Z_{-1},
\end{equation}
where $\hat{u}_{+1}$ and $\hat{u}_{-1}$ are the slowly varying complex envelope functions, $Z_{l}= e^{il(kz -\om t)}$
is the fast oscillating function, $\omega$ and $k$ are the frequency and the wave number of the carrier wave.
Because $u$ and $u^{*}$ are complex conjugation functions, therefore for the reality of $U$ we set:
$ \hat{u}_{+1}= \hat{u}^{*}_{-1}$.

Substituting Eq.(4) into (3) the real function $U(z, t)$  can be represented in the form
\begin{equation}\label{uU1}
 U=\sum_{l=\pm1}\hat{u}_{l}Z_{l}.
\end{equation}

The purpose of the present work is to consider the two-component vector breather solution of the function $U(z,t)$ using generalized perturbative reduction method Eq.(10), when the function $u(z,t)$ satisfies the Hirota equation (1) and the slowly varying envelope approximation, Eqs.(4) and (5).

The rest of this paper is organized as follows: Section II is devoted to the linear part of the Hirota equation for slowly varying complex envelope functions. In Section III, using the generalized perturbation reduction method, we will transform Eq.(9) to the coupled nonlinear  Schr\"odinger equations for auxiliary functions.
In Section IV, will be presented the explicit analytical expressions for the shape and parameters of the two-component nonlinear pulse. Finally, in Section V, we will discuss the obtained results.

\vskip+0.5cm
\section{The linear part of the Hirota equation}

The linear part of the Hirota equation (1) have the form
\begin{equation}\label{hiilin}
 i  \frac{\pa u}{\pa t} + \rho \frac{\pa^{2} u}{\pa z^{2}}  +i\sigma \frac{\pa^{3} u}{\pa z^{3}} =0.
\end{equation}

We consider a pulse whose duration satisfies  the condition $T >>\omega^{-1}$.
Substituting  Eq.(4) into (6) we obtain the dispersion relation
\begin{equation}\label{dis}
 \om =\rho k^{2} -\sigma k^3
\end{equation}
and the rest part of the linear equation
\begin{equation}\label{hirlin}
 (\pm i\frac{\pa \hat{u}_{\pm 1}}{\pa t}   \pm i A \frac{\pa \hat{u}_{\pm1}}{\pa z}
  +B\frac{\pa^{2} \hat{u}_{\pm1}}{\pa z^2} \pm i\sigma \frac{\pa^{3} \hat{u}_{\pm1}}{\pa z^3} )  Z_{\pm1} =0.
\end{equation}
where 
$$
A=2\rho k  -3 \sigma  k^{2},\;\;\;\;\;\;\;\;\;\;\;\;\;B= \rho -3\sigma k.
$$
We can rewrite Eq.(8) to the following form
\begin{equation}\label{lin2}
 (l i\frac{\pa \hat{u}_{l}}{\pa t}  +l i A \frac{\pa \hat{u}_{l}}{\pa z}   +B \frac{\pa^{2} \hat{u}_{l}}{\pa z^2}
+l i\sigma \frac{\pa^{3} \hat{u}_{l}}{\pa z^3} )  Z_{l} =0,\;\;\;\;\;\;\;\;\;\;\;\;\;\;\;\;\;\;\;l=\pm 1.
\end{equation}

\vskip+0.5cm

\section{The generalized perturbative reduction method}

For the investigation of the two-component nonlinear solitary wave solution of Eqs.(1) and (5) we apply the generalized perturbative reduction method [15-23] by means of which we can transform  the Hirota equation into the coupled nonlinear Schr\"odinger equations for auxiliary functions. In this method the function $\hat{u}_{l}(z,t)$ can be represented as:
\begin{equation}\label{cemi}
\hat{u}_{l}(z,t)=\sum_{\alpha=1}^{\infty}\sum_{n=-\infty}^{+\infty}\varepsilon^\alpha Y_{l,n} f_{l,n}^ {(\alpha)}(\zeta,\tau),
\end{equation}
where
$$
Y_{l,n}=e^{in(Q_{l,n}z-\Omega_{l,n} t)},\;\;\;\;\;\;\;\;\;\;\;\;\;\;\zeta_{l,n}=\varepsilon Q_{l,n}(z-{v_{g;}}_{l,n} t),
$$$$
\;\;\;\tau=\varepsilon^2 t,\;\;\;\;\;\;\;\;\;\;\;\;\;\;\;\;\; {v_{g;}}_{l,n}=\frac{\partial \Omega_{l,n}}{\partial  Q_{l,n}},
$$
$\varepsilon$ is a small parameter. Such an expansion allows us to separate from $\hat{u}_{l}$ the even more slowly changing auxiliary
function $ f_{l,n}^{(\alpha )}$. It is assumed that the quantities $\Omega_{l,n}$, $Q_{l,n}$, and $f_{l,n}^{(\alpha)}$ satisfy the conditions:
\begin{equation}\label{rpt}\nonumber
\omega\gg \Omega_{l,n},\;\;\;\;\;\;\;\;k\gg Q_{l,n},
\end{equation}
$$
\left|\frac{\partial f_{l,n}^{(\alpha )}}{
\partial t}\right|\ll \Omega_{l,n} \left|f_{l,n}^{(\alpha )}\right|,\;\;\;\;\;\;\;\;\left|\frac{\partial
f_{l,n}^{(\alpha )}}{\partial z }\right|\ll Q_{l,n}\left|f_{l,n}^{(\alpha )}\right|,
$$
for any value of indexes $l$ and $n$.

Substituting Eq.(10) into (9) we obtain the linear part of the Hirota equation in the form
\begin{equation}\label{hhj}
  \sum_{l=\pm 1}\sum_{\alpha=1}^{\infty}\sum_{n=-\infty}^{+\infty}\varepsilon^{\alpha}  Z_{l} Y_{l,n}
[ W_{l,n} +i\varepsilon J_{l,n} \frac{\partial }{\partial \zeta}+l i\varepsilon^2 \frac{\partial }{\partial \tau}
  + \varepsilon^{2} H_{l,n}  \frac{\partial^{2} }{\partial \zeta^{2}} +  O(\varepsilon^3))] f_{l,n}^{(\alpha)}  =0,
\end{equation}
where
\begin{equation}\label{iun}
W_{l,n}= l n\Omega_{l,n}  - l n A Q_{l,n}   - B Q^{2}_{l,n} + l   n \sigma Q^{3}_{l,n},
$$$$
J_{l,n}=-l  Q_{l,n}  {v_{g;}}_{l,n} + l  A  Q_{l,n} + 2B  n Q^{2}_{l,n}  -  3 l \sigma   Q^{3}_{l,n},
$$$$
H_{l,n}= Q^{2}_{l,n} (B -  3 ln  \sigma  Q_{l,n}).
\end{equation}

Eq.(11) contains four independent equations for different values of the indexes $l=\pm1$ and $n=\pm1$.

If we equating to zero, the terms of the Eq.(11) with the same powers of $\varepsilon$, we will be able to a several of equations. In the first order of $\varepsilon$ we have the equation
\begin{equation}\label{fr}
 \sum_{l=\pm 1}\sum_{n=-\infty}^{+\infty} Z_{l} Y_{l,n} W_{l,n} f_{l,n}^{(1)}=0.
\end{equation}
From the equations (12) we can see that following equations
\begin{equation}\label{5r}
W_{+1,+1}=W_{-1,-1},\;\;\;\;\;\;\;\;\;\;\;\;\;\;\;\;\;W_{+1,-1}=W_{-1,+1}
\end{equation}
are valid.

Taking into account Eqs.(13) and (14), from Eq.(12), for $W_{l,n}$, we obtain four independent equations for the connection between the parameters $\Omega_{l,n}$ and $Q_{l,n}$.
But these four equations are reduced to the  following two independent equations:
\begin{equation}\label{diss1}
 \Omega_{\pm1,\pm1}  - A Q_{\pm1,\pm1}   - B  Q^{2}_{\pm1,\pm1} +  \sigma Q^{3}_{\pm1,\pm1}=0,
\end{equation}
when $ f_{\pm1,\pm1}^{(1)}\neq 0$, and
\begin{equation}\label{diss2}
 \Omega_{\pm 1,\mp 1}  - A Q_{\pm 1,\mp 1}    + B Q^{2}_{\pm 1,\mp 1} +  \sigma Q^{3}_{\pm 1,\mp 1}=0,
\end{equation}
when  $ f_{\pm1,\mp 1}^{(1)}\neq 0$.

From Eqs.(15) and (16) we obtain the expressions
\begin{equation}\label{v1}
{v_{g;}}_{\pm 1,\pm 1}= \frac{\partial \Omega_{\pm 1,\pm 1}}{\partial Q_{\pm 1,\pm 1}}=A+2BQ_{\pm 1,\pm 1}-3\sigma Q^{2}_{\pm 1,\pm 1},
$$$$
{v_{g;}}_{\pm 1,\mp 1}= \frac{\partial \Omega_{\pm 1,\mp 1}}{\partial Q_{\pm 1,\mp 1}}=A-2BQ_{\pm 1,\mp 1}-3\sigma Q^{2}_{\pm 1,\mp 1}.
\end{equation}

Taking into account Eqs.(12), (15) and (16), in the second order of $\varepsilon$, we obtain
$$
J_{\pm1,\pm1}=J_{\pm1,\mp1}=0,\;\;\;\;\;\;\;\;
 f_{+1,\pm2}^{(2)}=f_{-1,\pm2}^{(2)}=0.
$$

Next we consider the Hirota equation in the third order of $\varepsilon$, for which we have to take into account the nonlinear term of the Hirota equation (1).
Substituting Eq.(10) into (1) we obtain nonlinear part of the Hirota equation in the following form
\begin{equation}\label{nonlin}
\delta |u|^{2} u   +i 3 \alpha |u|^{2} u_{z} =
$$$$
Z_{+1}Y_{+1,+1} ( [ \delta   - 3 \alpha  (k + Q_{+1,+1}) ] |f_{+1,+1}^ {(1)}|^{2} +[ 2 \delta  - 3 \alpha  (2k + Q_{+1,+1}  -  Q_{+1,-1}) ] |f_{+1,-1}^ {(1)}|^{2}) f_{+1,+1}^ {(1)}
$$$$
+ Z_{+1}Y_{+1,-1} ([\delta  - 3 \alpha (k -  Q_{+1,-1})] |f_{+1,-1}^ {(1)}|^{2} + [2 \delta - 3 \alpha (2k + Q_{+1,+1}  -  Q_{+1,-1})] |f_{+1,+1}^ {(1)}|^{2}) f_{+1,-1}^ {(1)}.
\end{equation}

Combining Eqs.(1), (11) and (18) we obtain the system of nonlinear equations
\begin{equation}\label{H}
Z_{+1} Y_{+1,+1} [i \frac{\partial  f_{+1,+1}^{(1)} }{\partial \tau}  +  H_{+1,+1} \frac{\partial^{2} f_{+1,+1}^{(1)}}{\partial \zeta^{2}_{+1,+1}}
$$$$
+ ([\delta - 3\alpha (k + Q_{+1,+1})] |f_{+1,+1}^ {(1)}|^{2} +[2 \delta - 3 \alpha (2k + Q_{+1,+1} - Q_{+1,-1})] |f_{+1,-1}^ {(1)}|^{2}) f_{+1,+1}^ {(1)}]=0,
$$$$
Z_{+1} Y_{+1,-1} [ i \frac{\partial  f_{+1,-1}^{(1)} }{\partial \tau}  +  H_{+1,-1}  \frac{\partial^{2} f_{+1,-1}^{(1)}}{\partial \zeta^{2}_{+1,-1}}  +$$$$
([\delta - 3 \alpha (k - Q_{+1,-1})] |f_{+1,-1}^ {(1)}|^{2} + [2 \delta - 3 \alpha (2k + Q_{+1,+1} - Q_{+1,-1})] |f_{+1,+1}^ {(1)}|^{2}) f_{+1,-1}^ {(1)}]=0.
\end{equation}

\vskip+0.5cm
\section{The two-component vector breather of the Hirota equation}

After transformation back to the variables $z$ and $t$, from Eqs.(19) we obtain the coupled nonlinear Schr\"odinger equations for auxiliary
functions $\lambda_{\pm}=\varepsilon  f_{+1,\pm 1}^{(1)}$ in the form
\begin{equation}\label{pp2}
i (\frac{\partial \lambda_{\pm}}{\partial t}+v_{\pm} \frac{\partial  \lambda_{\pm}} {\partial z}) + p_{\pm} \frac{\partial^{2} \lambda_{\pm} }{\partial z^{2}}
+q_{\pm}| \lambda_{\pm}|^{2}\lambda_{\pm} +r_{\pm} |\lambda_{\mp}|^{2} \lambda_{\pm}=0,
\end{equation}
where
\begin{equation}\label{Hnmin}
v_{\pm}={v_{g;}}_{+ 1,\pm 1}=2\rho (k \pm  Q_{\pm}) \mp 6\sigma k Q_{\pm}-3 \sigma ( k^2 + Q^{2}_{\pm }),
$$$$
p_{\pm}=\rho -3\sigma ( k  \pm    Q_{\pm}),
$$$$
q_{\pm}=\delta - 3\alpha (k \pm  Q_{\pm}),
$$$$
r_{\pm}=2 \delta - 3 \alpha (2k + Q_{+} - Q_{-}).
\end{equation}
\begin{equation}\label{OmQ}
\Omega_{+1,+1}= \Omega_{-1,-1}= \Omega_{+},\;\;\;\;\;\;\;\;\;\;\;\;\;\;\;\;\;\;\;\;\;\;\;\;\;  \Omega_{+1,-1}= \Omega_{-1,+1}= \Omega_{-},
$$$$
Q_{+1,+1}= Q_{-1,-1}= Q_{+},\;\;\;\;\;\;\;\;\;\;\;\;\;\;\;\;\;\;\;\;\;\;\;\;\; Q_{+1,-1}= Q_{-1,+1}= Q_{-}.
\end{equation}

Using Eq.(22), we can combine the equations (15) and (16) together to present these equation in the more compact form
\begin{equation}\label{dis2}
 \Omega_{\pm}  - k (2\rho   -3 \sigma  k) Q_{\pm}   \mp  (\rho -3\sigma k) Q^{2}_{\pm} +  \sigma Q^{3}_{\pm}=0.
\end{equation}

The solution of Eq.(20) we seek in the form of [21,22]
\begin{equation}\label{ue1}
\lambda_{\pm }=\frac{A_{\pm }}{b T}Sech(\frac{t-\frac{z}{V_{0}}}{T}) e^{i(k_{\pm } z - \omega_{\pm } t )},
\end{equation}
where $A_{\pm },\; k_{\pm }$ and $\omega_{\pm }$ are the real constants, $V_{0}$ is the velocity of the nonlinear wave. We assume that
\begin{equation}\label{kom}
k_{\pm }<<Q_{\pm },\;\;\;\;\;\;\omega_{\pm }<<\Omega_{\pm }.
\end{equation}

The connections between  $A_{\pm },\; k_{\pm }$ and $\omega_{\pm }$  are given by
\begin{equation}\label{ttw}
A_{+}^{2}=\frac{p_{+}q_{-}- p_{-}r_{+}}{p_{-}q_{+}- p_{+}r_{-}}A_{-}^{2},
$$$$
k_{\pm }=\frac{V_{0}-v_{\pm}}{2p_{\pm}},
$$$$
\omega_{+}=\frac{p_{+}}{p_{-}}\omega_{-}+\frac{V^{2}_{0}(p_{-}^{2}-p_{+}^{2})+v_{-}^{2}p_{+}^{2}-v_{+}^{2}p_{-}^{2}
}{4p_{+}p_{-}^{2}}.
\end{equation}

Substituting Eq.(24) into (20), when the complex envelope function $u$ satisfies the Hirota equation (1), and taking into account Eqs.(4), (5) and (10), we obtain the two-component pulse for the function $U$:
\begin{equation}\label{vb}
U(z,t)= \frac{2}{b T}Sech(\frac{t-\frac{z}{V}}{T})\sum_{j=\pm} A_{j} \cos[(k+j Q_{j}+k_{j})z -(\om +j \Omega_{j}+\omega_{j}) t],
\end{equation}
where $T$ is the width of the two-component nonlinear pulse,
\begin{equation}\label{tb}
\frac{1}{(bT)^2}=2 \; \frac{v_{+}k_{+}+k_{+}^{2}p_{+}-\omega_{+}}{A_{+}^{2}q _{+} + A_{-}^{2}r _{+}}.
\end{equation}

This pulse oscillating with the sum $\omega+\Omega_{+}$
and difference $\omega-\Omega_{-}$ of the frequencies and the wave numbers $k+Q_{+}$ and $k-Q_{-}$ (taking into account Eq.(25)).

\vskip+0.5cm
\section{Conclusion}

In the present paper, we study the two-component nonlinear solitary wave (vector breather) which can be describe by means of the Hirota equation (1). Using the generalized perturbation reduction method Eq.(10), the Hirota equation is transformed to the coupled nonlinear Schr\"odinger equations for auxiliary functions $\lambda_{\pm }$, Eq.(20).
We investigate nonlinear wave with the width $T>>\Omega_{\pm }^{-1}>>\omega^{-1}$ and is shown that  under such condition can be formed two-component vector breather, the one component of which oscillate with the sum $\omega+\Omega_{+}$ of the frequencies and the wave numbers $k+Q_{+}$  and the second component oscillate with the  difference $\omega-\Omega_{-}$ of the frequencies and the wave numbers  $k-Q_{-}$ in the region of the carrier wave frequency $\omega$ and wave number $k$. These wave components propagate in the same direction, with the identical velocities and have the same polarizations. The explicit analytical expression of the shape and parameters of the nonlinear pulse by Eqs.(17), (21), (26), (27) and (28), are presented. The dispersion relation and the connection between the parameters $\Omega_{\pm}$ and $Q_{\pm}$ are determined from Eqs.(7) and (23).

We have to note that the similar two-component nonlinear waves we met in the different fields of research for various nature of waves: optical, acoustic, magnetic, hydrodynamics and others [5, 18, 24-27]. In the theory of the self-induced transparency such wave is called the vector $0\pi$ pulse [20-22].

\vskip+0.5cm

\end{document}